\documentclass[12pt]{article}
\newcommand\beq{\begin{eqnarray}}
\newcommand\eeq{\end{eqnarray}}
\newcommand\me{\frac{m_{\nu}}{E}}
\newcommand\qove{\frac{q}{E}}
\newcommand\qovm{\frac{q}{2M}}
\newcommand\qom{\frac{q}{M}}
\newcommand\fip{\frac{\mid \phi_{\mu}(0) \mid^{2}}{4\pi}}
\newcommand\fipp{\frac{\mid \phi_{\mu}(0) \mid^{2}}{8\pi}}
\newcommand\vcl{\mid C_{V}^{L} \mid^{2}}
\newcommand\acl{\mid C_{A}^{L} \mid^{2}}
\newcommand\vcr{\mid C_{V}^{R} \mid^{2}}
\newcommand\acr{\mid C_{A}^{R} \mid^{2}}
\newcommand\vlal{C_{V}^{L}C_{A}^{L*}}
\newcommand\vlar{C_{V}^{L}C_{A}^{R*}}
\newcommand\vral{C_{V}^{R}C_{A}^{L*}}
\newcommand\alar{C_{A}^{L}C_{A}^{R*}}
\newcommand\vlvr{C_{V}^{L}C_{V}^{R*}}

\newcommand\vrar{C_{V}^{R}C_{A}^{R*}}
\newcommand\pmo{|\vec{P_{\mu}}|}
\begin{document}
\title{RIGHT-HANDED VECTOR $V$ AND AXIAL $A$ COUPLINGS IN WEAK
INTERACTIONS}
\author{Wies\l{}aw Sobk\'ow \\
Institute of Theoretical Physics, University of Wroc\l{}aw,\\
Pl. M. Born 9, PL-50-204~Wroc\l{}aw, Poland\\
e-mail: {\tt Sobkow@rose.ift.uni.wroc.pl}}
\maketitle

\begin{abstract}
In this paper a scenario admitting the participation  of the
right-handed
vector $ V_{R}$ and axial $A_{R}$ couplings with the conservation of
the left-handed
standard $(V, A)_{L}$
couplings  is considered. The research is based on the muon capture by proton.
We consider muon capture at the level of the Fermi theory, whose hamiltonian
describes the four-fermion point (contact) interaction.
Neutrinos are assumed to be massive  and to be  Dirac fermions. We
propose  neutrino observables, it means transverse components of
the neutrino polarization, both $ T$-odd and $ T$-even. That would
be a test verifying the participation of the $(V, A)_{R}$
couplings in muon capture. The measurements of  nuclear
observables and of  longitudinal neutrino polarization do not
offer such possibilities because  of the suppressing of
interferences between  the $(V, A)_{L}$ and $(V, A)_{R}$ couplings
caused by the neutrino mass.
Using the current data from
$\mu$-decay and inverse $\mu$-decay, the magnitude of effects
coming from the transverse components of the neutrino polarization
can be determined. Our considerations are model-independent. We
give the lower bound of $305\; GeV$ on the mass of the
right-handed gauge boson. This limit is compatible with the
current bounds on the mass of the $W_{R}$ received from the weak
interaction processes at low energy.

\end{abstract}

\section{Introduction}

The  present theory of weak interactions (the Standard Model of
electroweak interactions \cite{Glashow,Wein,Salam}) describes only
what has been measured so far.  These are, most of all, the
measurements of nuclear observables and of observables for massive
leptons. It means the measurement of the electron helicity
\cite{Bob}, the indirect measurement of the neutrino helicity
\cite{Gold}, the asymmetry in the distribution of the electrons
from $\beta$-decay \cite{CWu}, the experiment with muon decay
confirming parity violation \cite{Gar}. Basing their inferences on
these results, among others, Feynman, Gell-Mann and Sudarshan,
Marshak \cite{Gell,Sud} established  that only left-handed vector
$ V$, axial $ A$ couplings are involved in weak interactions
because this yields the maximum symmetry breaking under space
inversion, under charge conjugation; the two-component neutrino
theory of negative helicity;  the conservation of the combined
symmetry $ CP$ and of the lepton number. The Fermi hamiltonian,
being low-energy approximation of the Salam-Weinberg model, has
thus a vector-axial $(V-A)$ structure, and the three remaining
scalar $ S$, tensor $ T$, pseudoscalar $ P$ couplings are
eliminated by the assumption that only left-handed states can take
part. The $ V-A$ theory has a chiral symmetry.

The investigation  of the completeness of the Lorentz structure
and of the handedness structure of weak interactions  at low
energies can be reduced to two main scenarios. The first
conception assumes the participation of the scalar $ S$, tensor $
T$ and pseudoscalar $ P$ couplings in addition to the standard
vector $ V$ and axial $ A$ couplings. S. C. Wu \cite{SWu}
indicated explicitly that possibility. According to her, both
left-handed $(V, A)_{L}$ couplings and exotic right-handed $(S, T,
P)_{R}$ couplings may be responsible for the negative electron
helicity observed in $\beta$-decay. K. Mursula {\sl et al.}
\cite{Murs} analyzed all the available data on the charged
leptonic weak interactions while testing different models which
admit the participation of additional $ (S, T, P)$ couplings
beside the standard $ (V, A)$ couplings.
\par The other conception assumes that the right-handed vector $V_{R}$ and axial
$A_{R}$ couplings participate in weak interactions beside
left-handed standard $ (V, A)_{L}$ couplings. This scenario is
studied and analyzed in this work. There are many theoretical and
experimental papers devoted to this problem. The models with
$SU(2)_{L}\times SU(2)_{R}\times U(1)$ as the gauge group emerged
first in the framework of a class of grand unified theories (GUT)
\cite{Pati}. The manifest left-right symmetry model \cite{Beg}
predicts the existence of the additional heavy vector bosons of
the masses much larger than the masses of the  bosons of the
Standard Model. B\'eg considered the bounds on the admixtures of
the right-handed currents obtained from the measurements of the
lepton polarization in semileptonic decays and by the
determination of the parameters characterizing  the spectrum in
muon decay. Herczeg model is the generalization of the manifest
left-right symmetric model, in which the fermions couple to
distinct charged gauge-boson fields $W_{L}$ and $W_{R}$ with the
different coupling constants $g_{L}$ and $g_{R}$, respectively
\cite{Her}. The effective hamiltonian has the structure of the
four-fermion point interaction, for both muon decay and
semileptonic processes. There are many experimental constraints on
the possible mass of the right-handed vector bosons $W_{R}$
obtained from weak interaction processes at low energy and from
high energy collider experiments. The lower mass limits for the
$W_{R}$ received at the Tevatron collider are $M_{R}\geq 652\;
GeV\;  (95\% \;CL$, CDF-collaboration \cite{Abe}) and $M_{R}\geq
720 \;GeV\; (95\% \; CL$, D0-collaboration \cite{Abachi}),
respectively. The lower bound obtained from $K_{0} -
\overline{K_{0}}$  mass difference is $M_{R}\geq 1.6 \;TeV$
\cite{Beall}.
 A. Jodidio {\sl et al.} \cite{Jodid} measured the positron spectrum from
muon decay, which allowed them to give the lower bound of $ 432\;
GeV \; (90 \% \;CL)$ on the possible mass of a new vector boson.
J. Maalampi {\sl et al.} \cite{Maalampi} explored the structure of
the charged leptonic weak currents in the framework of the
$SU(2)_{L}\times SU(2)_{R}\times U(1)$ models. They fitted the
parameters of this model to experimental results obtained from
pseudoscalar mezon decay, muon decay, nuclear $\beta$-decay and
inverse muon decay. It allowed them to determine the values of the
mass ratio of the charged gauge bosons and the mixing angle. N. B.
Shul'gina \cite{Szul} introduced the admixtures of the
right-handed $ (V, A)_{R}$ currents into the interaction
lagrangian, which helped to explain, e. g., the neutron paradox.
Recent measurements of the longitudinal polarization of the
positrons emitted by the polarized $ ^{107}In$ and $ ^{12}N$
nuclei gave the lower limit of $ 306\; GeV\; (90 \% \; CL)$ on the
mass of the right-handed gauge boson \cite{Sever}. The recommended
lower constraint on the mass of the additional gauge boson is
$M_{R}>549\; GeV$ \cite{Data}. M. Zra\l{}ek {\sl et al.}
\cite{Zralek} considered the possibility of the existence of
neutrino magnetic moments in the framework of the left-right
symmetry models. That could be especially interesting in the
context of the solar neutrino deficit. However, all these limits
are {\sl model-dependent} and they can be considerably weakened.
The stringent bound $M_{R}\geq 1.6 \;TeV$ can be relaxed to the
$300 \;GeV$ range \cite{Olness,Lang} if one assumes that the
Cabbibo-Kobayashi-Maskawa matrix elements for the right-handed
quarks and for the left-handed quarks are not identical, and also
that the $SU(2)_{L,R}$ gauge coupling constants are distinct,
$g_{L}\not= g_{R}$. Therefore one should give the bounds for the
nonstandard $(V, A)_{R}$ couplings {\sl without model assumptions}
\cite{Fetscher}. There are the present limits on all possible
coupling constants obtained from normal muon decay and inverse
muon decay \cite{Data}.

However, to verify uniquely a scenario admitting the possible
participation of the right-handed $(V, A)_{R}$ couplings beside
the standard  $(V, A)_{L} $ couplings in weak interactions at low
energies, one proposes {\sl neutrino observables} in the process
of $\mu$-capture by proton. Both $ T$-odd and $ T$-even transverse
components of the neutrino polarization are taken into account.
Only in the quantities of this type the interference terms between
the standard $(V, A)_{L}$ and nonstandard $(V, A)_{R}$ couplings
appear and they do not depend {\sl explicite} on the neutrino
mass. It is analogical to the situation analysed by T.D. Lee and
C.N. Yang in $\beta$-decay \cite{Lee}. They proposed an observable
which was pseudoscalar under space inversion to determine uniquely
if parity is violated. Only in this quantity interferences between
couplings for the parity-conserving interactions $C$ and  the
parity-nonconserving interactions $C'$ can appear. Using the
current data from $\mu$-decay and inverse $\mu$-decay,  the
magnitude of effects coming from the transverse components of the
neutrino polarization can be determined. At the end, we can derive
the lower limit on the mass of the right-handed gauge bosons
(assuming, for example, manifest left-right symmetry).

 Recently J. Sromicki
\cite{Sromicki} measured the $CP$-odd transverse electron
polarization in $^{8}Li$ $\beta$-decay. The final results
indicated the compatibility with the Standard Model prediction and
$CP$-conservation in $\beta$-decay. Armbruster et al. \cite{Armbr}
measured the energy spectrum of electron neutrinos $\nu_{e}$ from
$\mu$-decay at rest in the KARMEN experiment using the reaction
$^{12}C(\nu_{e},e^{-})^{12}N_{g.s.}$. They determined the upper
limit of $|g_{RL}^{S}+2g_{RL}^{T}|\leq 0.78 \;(90\% \;CL)$ on the
possible interference term between scalar $S$ and tensor $T$
couplings. M. Abe et al. \cite{MAbe} searched for T-odd transverse
components of the muon polarization in $K^{+}\rightarrow
\pi^{0}+\mu^{+}+\nu_{\mu}$ decay at rest. They pointed out that
the contribution to this observable from the Standard Model is of
the order of $10^{-7}$, so nonzero values of this quantity would
indicate the beginning of new physics beyond the Standard Model.
In our case, there is no contribution to the transverse neutrino
polarization from the Standard Model (massless Dirac neutrinos),
so nonzero values of such observable would be the unique proof of
the participation of the $(V, A)_{R}$ couplings and of the
production of the right-handed neutrinos. These last experiments
made at high precision show that the problem of the completeness
of the Lorentz structure and of the handedness structure of weak
interactions is  still explored.

The purpose of this paper is motivated by the desire to test how
right-handed vector $ V_{R}$ and axial $ A_{R}$ couplings with the
participation of left-handed standard $(V, A)_{L}$  couplings
enter different observables such as: longitudinal and  transverse
neutron polarization, longitudinal neutrino polarization and, most
of all, transverse neutrino polarization.

The structure  of the work is as follows: Sect.\  2 concentrates
on the qualitative description of muon capture and on the
assumptions concerning the calculations. In Sect.\  3 the results
obtained for the transverse, longitudinal neutron polarization and
longitudinal neutrino polarization, among others, are presented.
In Sect.\  4 the results for the transverse neutrino polarization
are dealt with. Sect.\  5 gives the conclusions. In these
considerations the system of natural units with $\hbar=c=1$, Dirac
hermitian matrices $\gamma_{\lambda}$ and the four-plus metric
are used \cite{Morit}.

\section{Muon capture by proton}
\par
The research is based on the reaction of the muon capture by proton
$\mu^{-} + p \rightarrow n + \nu_{\mu} \label{mion} $. In the Standard
Model it is a coherent low-energy process at the lepton-quark level.
The typical energy transfer is of the order of $1 MeV$ and therefore
the space-time area within the interactions coincides with the size of
the muonic atom, because of that both hadrons and leptons participating
in this process are point objects. In the light of the above  muon
capture is considered at the level of the Fermi theory, whose
hamiltonian describes the local, derivative-free,
lepton-number-conserving, four-fermion point (contact) interaction.
Right-handed $(V, A)_{R}$ couplings are assumed to take part in muon
capture in addition to left-handed standard $(V, A)_{L}$ couplings. The
coupling constants are denoted as $C_{V}^{L} $,  $C_{A}^{L} $ and
$C_{V}^{R}$, $C_{A}^{R}$ respectively to the neutrino handedness.
\beq
H_{\mu^-} \label{hamilt1} & = & C_{V}^{L}({\overline\Psi}_{\nu}
    \gamma_{\lambda}(1 + \gamma_{5})\Psi_{\mu})
    ({\overline\Psi}_{n}\gamma_{\lambda}\Psi_{p}) \\
&   & \mbox{} + C_{A}^{L}({\overline\Psi}_{\nu}
    i\gamma_{5}\gamma_{\lambda}(1 + \gamma_{5})\Psi_{\mu})
    ({\overline\Psi}_{n}i\gamma_{5}\gamma_{\lambda}\Psi_{p})    \nonumber \\
&   & \mbox{} + C_{V}^{R}({\overline\Psi}_{\nu}
    \gamma_{\lambda}(1 - \gamma_{5})\Psi_{\mu})
    ({\overline\Psi}_{n}\gamma_{\lambda}\Psi_{p})  \nonumber\\
&   & \mbox{} + C_{A}^{R}({\overline\Psi}_{\nu}
    i\gamma_{5}\gamma_{\lambda}(1 - \gamma_{5})\Psi_{\mu})
    ({\overline\Psi}_{n}i\gamma_{5}\gamma_{\lambda}\Psi_{p})  \nonumber
\eeq
    where $\Psi_{\mu}, \Psi_{\nu}, \Psi_{p}, \Psi_{n}$ -
Dirac bispinors for the muon, muonic neutrino, proton and neutron.
The above hamiltonian can be derived from the one by T. D. Lee and
C. N. Yang \cite{Lee}, when the following expression is put $
C_{V} + C_{V'}\gamma_{5} = C_{V}^{L}(1 + \gamma_{5}) + C_{V}^{R}(1
- \gamma_{5}), \; C_{A} + C_{A'}\gamma_{5} = C_{A}^{L} (1 +
\gamma_{5}) + C_{A}^{R}(1 - \gamma_{5})$, where $C_{V}^{L}=(C_{V}
+ C_{V'})/2,\; C_{V}^{R}=(C_{V} - C_{V'})/2,\; C_{A}^{L}=(C_{A} +
C_{A'})/2, \; C_{A}^{R}=(C_{A} - C_{A'})/2, \; C_{V}, C_{A},
C_{V'}, C_{A'}$ - the vector and axial couplings for the
parity-conserving interactions and the parity-nonconserving
interactions. The remaining couplings are omitted
$C_{S}=C_{S'}=C_{P}=C_{P'}=C_{T}=C_{T'}=0$. The Fermi hamiltonian,
Eq. (\ref{hamilt1}), can be modified if one puts $C_{V}^{L} -
C_{A}^{L}\gamma_{5} = (C_{V}^{L} - C_{A}^{L})(1 + \gamma_{5})/2 +
(C_{V}^{L} + C_{A}^{L})(1 - \gamma_{5})/2, \; C_{V}^{R} +
C_{A}^{R}\gamma_{5} = (C_{V}^{R} + C_{A}^{R})(1 + \gamma_{5})/2 +
(C_{V}^{R} - C_{A}^{R})(1 - \gamma_{5})/2$, and then, one obtains
the effective hamiltonian of the form:
\beq H_{\mu^-} \label{hamilt2}
& = & \frac{C_{V}^{L} - C_{A}^{L}}{2}({\overline\Psi}_{\nu}
    \gamma_{\lambda}(1 + \gamma_{5})\Psi_{\mu})
    ({\overline\Psi}_{n}\gamma_{\lambda}(1 + \gamma_{5})\Psi_{p}) \\
&   &  \mbox{} + \frac{C_{V}^{L} + C_{A}^{L}}{2}({\overline\Psi}_{\nu}
    \gamma_{\lambda}(1 + \gamma_{5})\Psi_{\mu})
    ({\overline\Psi}_{n}\gamma_{\lambda}(1 - \gamma_{5})\Psi_{p})\nonumber\\
&   & \mbox{} + \frac{C_{V}^{R} + C_{A}^{R}}{2}({\overline\Psi}_{\nu}
    \gamma_{\lambda}(1 - \gamma_{5})\Psi_{\mu})
    ({\overline\Psi}_{n}\gamma_{\lambda}(1 + \gamma_{5})\Psi_{p})\nonumber\\
&   & \mbox{} + \frac{C_{V}^{R} - C_{A}^{R}}{2}({\overline\Psi}_{\nu}
    \gamma_{\lambda}(1 - \gamma_{5})\Psi_{\mu})
    ({\overline\Psi}_{n}\gamma_{\lambda}(1 - \gamma_{5})\Psi_{p})  \nonumber
\eeq
Muonic neutrinos are assumed to be massive and of Dirac nature.
The nonrelativistic approximation is applied both for the nucleons and the muon
in the $1s$ state.
To describe muon capture the following observables are used:
$\vec{P}_{\mu} $ - the initial muon polarization in the $1s$ state,
$\vec{S}_{\nu}$ -  the operator of the neutrino spin,
$\vec{q}$ - the momentum of the outgoing neutrino,
$\vec{J}_{n} $ - the operator of the neutron spin.
 $\vec{P}_{\mu} $ and $\vec{q}$ are assumed to be
perpendicular to each
other. To illustrate the calculation method the definition of the $T$-even
components of the transverse neutrino polarization is given
$ <\vec{S}_{\nu} \cdot \hat{\vec{P}}_{\mu}>_{f}
\equiv   Tr[ \vec{S}_{\nu}\cdot \hat{\vec{P}}_{\mu} \rho_{f}]$,
where
$\rho_{f}$ - the density operator of the final state (neutrino-neutron),
$\hat{\vec{P}}_{\mu}$ - the direction of the muon polarization in the
$1s$ state. The calculations are made with the Reduce computer program.

\section{ Longitudinal
and transverse neutron non-po\-la\-ri\-za\-tion
and  longitudinal neutrino non-po\-la\-ri\-za\-tion}
\par
In this section the results for the longitudinal and transverse neutron
polarization and for the longitudinal neutrino polarization are presented.
The final results are as follows:
\beq
 < \hat{\vec{q}} \cdot \vec{J}_{n}>_{f}
& = & \fip \{\mbox{} \qove (\acl - \acr)   \\
&  & \mbox{} + Re[(2\qom + \qove) (\vrar - \vlal) \nonumber\\
&  & \mbox{} +  \me \qom(\vral - \vlar)] \} \nonumber
\eeq
\beq
< \vec{J}_{n}\cdot(\hat{\vec{P}_{\mu}} \times \hat{\vec{q}}) >_{f}
& = & \fip \pmo Im\{(\qove + \qom ) (\vrar - \vlal)  \\
&  & \mbox{}  +  \me \qom (\vral - \vlar - \alar - \vlvr) \} \nonumber
\eeq
\beq
 < \vec{S}_{\nu}\cdot \hat{\vec{q}}>_{f}
& = & \fip \{Re[\qom(\vlal - \vrar)  \\
&  & \mbox{} + \me \qom (\vlar - \vral)] \nonumber \\
&  & \mbox{} + (\frac{1}{2}\qove +\qovm)(\vcr - \vcl) \nonumber\\
&  & \mbox{} +  (\frac{3}{2}\qove + \qovm)(\acr - \acl)\} \nonumber
\eeq
where, $\phi_{\mu}(0)$ - the value of the large radial component  of the muon
Dirac bispinor for $r=0$, $\hat{\vec{q}}$ - the direction of the neutrino
momentum,  $\pmo$ - the value of the muon polarization in  the $1s$ state,
$q/2M$ - the momentum corrections, $ q, E, m_{\nu},
M$ - the value of the neutrino momentum, its energy, its mass and the
nucleon mass, respectively.

It can be noticed that in these observables the occurrence of the
interferences between the right-handed $(V, A)_{R}$ and left-handed
$(V, A)_{L}$   couplings depends {\sl explicite} on the muonic neutrino
mass. Thus, the so-called "conspiracy" of interference terms appears
here. This "conspiracy" makes the measurement of the relative phase
between these two coupling types impossible because  a very small mass
of the neutrino $(m_{\nu}< 0.17 \;MeV \;CL= 90\% $ \cite{Data}) at its
high energy $(E_{\nu}\simeq 100 \;MeV)$ suppresses such an interference
in practice. The additional interference attenuation is further caused
by the momentum corrections. The factor $(m_{\nu}/E)  (q/M) \simeq 17
\cdot 10^{-5}$ is very small.
We can see that new
interference contributions can not be detected at the present level
of experimental precision.
Longitudinal neutrino polarization behaves as a typical nuclear quantity.
The "conspiracy" of interference terms caused by the neutrino mass occurs
here.
The measurement of this observable would not allow the unique determination
of the possible participation of the $(V, A)_{R}$ couplings. Therefore, the
observables in which such difficulties do not appear are proposed.

\section{ Why transverse components of neutrino polarization?}
\par
In this section  the results for two $T$-odd neutrino observables and for one
$T$-even neutrino quantity are given. All these three cases concern the
transverse neutrino polarization. In practice, that would mean the
measurements of the components of the neutrino polarization perpendicular to
its direction of momentum. The final results  are as follows:
\beq \label{Wynik1}
\lefteqn{< \vec{S}_{\nu}\cdot(\hat{\vec{P}}_{\mu}\times \hat{\vec{q}}) >_{f} }\\
& = & \fip \pmo Im \{ (\qove + \qom)( \alar - \vlvr)\}  \nonumber
\eeq
\beq  \label{Wynik2}
< \vec{S}_{\nu}\cdot(\vec{J}_{n}\times \hat{\vec{q}}) >_{f}
& = & \fip Im \{ (\qove + \qom)(\vral - \vlar)  \\
&  & \null + \qom \vlvr + ( 2\qove + \qom )\alar \nonumber \\
&  & \null + \me\qom (\vrar - \vlal)\} \nonumber
\eeq
\beq \label{Wynik3}
< \vec{S}_{\nu}\cdot \hat{\vec{P}}_{\mu}>_{f}
& = & \fip  \pmo Re \{(1 + \qove \qom)(\vlvr - \alar) \\
&  & \mbox{} + \frac{1}{2} \me(\vcr - \acr + \vcl - \acl)\} \nonumber
\eeq
The obtained result, Eq. (\ref{Wynik1}), consists exclusively of the
interference terms between
the left-handed standard $(V, A)_{L}$ couplings and the right-handed
nonstandard $(V, A)_{R}$
couplings, whose occurrence does not depend {\sl explicite} on the muonic
neutrino mass. It can be understood as the interference between the neutrino
waves of negative and positive chirality.
It creates the possibility of measuring the relative
phase between the two coupling types.
In the next observables, Eq. (\ref{Wynik2}) and (\ref{Wynik3}),
we have the additional dependence on the neutrino mass
which occurs
only at the terms of the type: $\vcr, \acr, \vcl, \acl, \vrar,
\vlal$.
It gives a very small contribution
in the relation to the main one
coming from  the  interferences between the $(V, A)_{L}$ and $(V, A)_{R}$
couplings.
There is no
contribution to these observables from the Standard Model, in which neutrinos
are  massless.
\par Now, we will express our coupling
constants $C_{V,A}^{L,R}$ by Fetscher's couplings $g^{\gamma}_{\epsilon
\mu}$ \cite{Fetscher} assuming the universality of weak interactions.
The induced couplings generated by the dressing of hadrons are
neglected as their presence does not change qualitatively the
conclusions about transverse neutrino polarization. Here, $\gamma= S,
V, T$ indicates a scalar, vector, tensor interaction; $\epsilon, \mu=L,
R$ indicate the chirality of the electron or muon and the neutrino
chiralities are uniquely determined for given $\gamma, \epsilon, \mu$.
We get the following relations: $C_{V}^{L}=A(g_{LL}^{V} + g_{RL}^{V}),
\; -C_{A}^{L}=A(g_{LL}^{V} - g_{RL}^{V}), \; C_{V}^{R}=A(g_{LR}^{V} +
g_{RR}^{V}), \; C_{A}^{R}=A(g_{LR}^{V} - g_{RR}^{V})$, where $A\equiv
(4G_{F}/\sqrt{2})cos\theta_{c}, \;G_{F}=1.16639(1)\times
10^{-5}GeV^{-2}$ \cite{Data} is the Fermi coupling constant,
$\theta_{c}$ is the Cabbibo angle ($cos\theta_{c}=0.9740 \pm 0.0010$
\cite{Data}). We can derive the contributions coming from the
$C^{L,R}_{V,A}$ coupling constants in $\mu$-capture, using the current
data \cite{Data}: $|C_{V}^{L}|>0.850 A, \;|C_{A}^{L}|>1.070 A,
\;|C_{V}^{R}|<0.093 A, \;|C_{A}^{R}|<0.027 A$. From the above, we can
see that the effects of the  transverse neutrino polarization connected
with the right-handed $(V, A)_{R}$ couplings may be of the order of
$7\%$.
This value is  {\sl model-independent}.
In this way, we can give the  lower
limit of  $M_{R}\geq 305 \;GeV$ on the mass of the
right-handed vector boson $W_{R}$
(for manifest left-right
symmetry, $m_{\nu}=0$, without $W_{L}-W_{R}$ mixing).
It is compatible
with the current bounds on the mass of the $W_{R}$ received from
the weak interaction processes at low energy \cite{Data}.
\par When the neutrinos are massive and only standard $(V, A)_{L}$ couplings
participate in muon capture, the transverse components of the
neutrino polarization could be observed in both cases:
\begin{equation}\protect\label{neut4}
    < \vec{S}_{\nu}\cdot(\vec{J}_{n}\times\hat{\vec{q}}) >_{f}
     =  \mbox{} - \fip \me \qom Im(\vlal)
    \end{equation}
    \beq\label{neut5}
< \vec{S}_{\nu}\cdot\hat{\vec{P}}_{\mu}>_{f}
    & = & \mbox{} \fipp \pmo  \me( \vcl - \acl)
     \eeq
However, we can see that the eventual effect of the nonzero transverse
components of
the neutrino polarization connected with the neutrino mass would be much
weaker than the one coming from the nonstandard couplings.
In such a scenario, the observable determined by the Eq.
(\ref{Wynik1}) always equals zero, $<
\vec{S}_{\nu}\cdot(\hat{\vec{P}}_{\mu}\times \hat{\vec{q}})
>_{f}=0$.
Because the direct measurement of the transverse neutrino polarization
in $\mu^{-}$-capture by proton is very difficult now, one proposes to
use the muonic neutrino-electron elastic scattering $\nu_{\mu} +
e^{-}\rightarrow e^{-} + \nu_{\mu}$ to detect the effects of the
transverse neutrino polarization connected with the nonstandard couplings.
From the differential cross section for this process   the
possible  neutrino-electron correlations in terms of the longitudinal
and transverse neutrino polarization will  be seen \cite{John} (in
preparation). Observing the change in the distribution
of the electrons in relation to the  distribution with standard neutrino,
one would obtain a clear signal of the nonzero values of the transverse
neutrino polarization. Maalampi {\sl et al.} \cite{Maalampi} considered
the presence of muonic neutrinos of a given initial polarization in the
inverse muon decay.
In the future the experimental verification of the hypothesis
concerning the transverse components of the neutrino polarization could
be carried out by the Fermi laboratory.
Currently at Fermilab,  the BooNE experiment \cite{Louis,Boone,Miniboone}
(The Booster
Neutrino Experiment) with the intense neutrino source is designed to
search for the muonic neutrino oscillations, the mass difference, the
mixing angle, the CP violation in the lepton sector, the muon-neutrino
disappearance signal, the neutrino magnetic moment, and the helicity
structure of the weak neutral current.
This experiment will also look for the non-oscillation
neutrino physics using, among others, the neutrino-electron elastic
scattering. At Fermilab, essentially all of muon-neutrinos  come from
$\pi^{+}$-decay. However, on the quark level, in $\mu^{-}$-capture and
$\pi^{+}$-decay, there is the same semileptonic interaction. Therefore
 the conclusions regarding transverse neutrino polarization in
$\mu^{-}$-capture are also correct for the muon-neutrinos coming from
$\pi^{+}$-decay.

\section{ Conclusions}
 The  measurements of the $T$-odd, Eq. (\ref{Wynik1}), transverse
components of the neutrino polarization could verify  the possibility
of the right-handed $(V, A)_{R}$ couplings participation in weak
interactions. They would also be the proof of the production of the
right-handed neutrinos. As far as the $T$-odd components of the
transverse  neutrino polarization are concerned, one would also obtain
a proof of the symmetry breaking under time inversion $T (CP)$ in a
semileptonic process. The measurement of longitudinal neutrino
polarization does not offer such possibilities because  of the
suppressing of interferences between  the $(V, A)_{L}$ and $(V, A)_{R}$
couplings caused by the neutrino mass. In this way, that will always
lead to the compatibility with the Standard Model. The similar
regularity can be observed in nuclear observables: longitudinal,
transverse neutron polarization and also probability of muon capture,
and the quantities of only this type are measured today.\\ The BooNE
experiment, which is now being constructed, will be able to measure the
nonstandard neutrino-electron correlations using the $\nu_{\mu} +
e^{-}\rightarrow e^{-} + \nu_{\mu}$ process. This experiment will be
started in the year 2001.

\vspace{.2cm}
I am greatly indebted to  Prof.\  S.\ Ciechanowicz for
many  useful and helpful discussions  and his interest in my research.
I owe much to Prof.\  M.\  Zra\l{}ek for  interesting critical remarks.


\addcontentsline{toc}{section}{Bibliography}
\begin{thebibliography}{99}

\bibitem{Glashow} S. L. Glashow,
{\sl Nucl. Phys.} {\bf 22}, 579  (1961).
\bibitem{Wein} S. Weinberg,
{\sl Phys. Rev. Lett.} {\bf 19}, 1264 (1967).
\bibitem{Salam} A. Salam,
{\sl Elementary Particle Theory}, N. Svartholm (Almquist and Wiksells),
Stockholm 1969, p. 367.
\bibitem{Bob} H. Frauenfelder  et al.,
{\sl Phys. Rev.} {\bf 106}, 386  (1957).
\bibitem{Gold} M. Goldhaber et al.,
{\sl Phys. Rev.} {\bf 109}, 1015 (1958).
\bibitem{CWu} C.S. Wu  et al.,
{\sl  Phys. Rev.} {\bf 105}, 1413 (1957); {\bf 106}, 1361 (1957).
\bibitem{Gar} R.L. Garwin et al.,
{\sl Phys. Rev.} {\bf 105}, 1415 (1957).
\bibitem{Gell} R. P. Feynman, M. Gell-Mann,
{\sl Phys. Rev.} {\bf 109}, 193 (1958).
\bibitem{Sud} E. C. G. Sudarshan, R. E. Marshak,
{\sl Phys. Rev.} {\bf 109}, 1860 (1958).
\bibitem{SWu} C.S. Wu, S.A. Moszkowski,
{\sl Beta decay}, Wiley, New York 1966, p. 163.
\bibitem{Murs} K. Mursula  et al.,
{\sl Nucl. Phys.}  {\bf B 219}, 321 (1983).
\bibitem{Pati} J.C. Pati, A. Salam, {\sl Phys. Rev. Lett.} {\bf 31}, 661,
(1973); {\sl Phys. Rev. } {\bf  D 10}, 275 (1974); R.N. Mohapatra,
J.C. Pati, {\sl Phys. Rev.} {\bf D 11}, 566, 2558 (1974); G.
Senjanovi\'c, R.N. Mohapatra, {\sl Phys. Rev.} {\bf D 12} 1502
(1975); R.N. Mohapatra, R.E. Marshak, {Phys. Lett.} {\bf B 91} 222
(1980).
\bibitem{Beg} M. A. B. B\'eg et al.,
{\sl Phys. Rev. Lett.} {\bf  38}, 1252 (1977).
\bibitem{Her} P. Herczeg,
{\sl Phys. Rev.}  {\bf D 34}, 3449 (1986).
\bibitem{Abe} F. Abe et al.,
{\sl Phys. Rev. Lett.} {\bf 74}, 2900 (1995).
\bibitem{Abachi} S. Abachi et al.,
{\sl Phys. Rev. Lett.} {\bf 76}, 3271 (1996).
\bibitem{Beall} G. Beall et al.,
{\sl Phys. Rev. Lett.} {\bf 48}, 848 (1982).
\bibitem{Jodid} A. Jodidio et al.,
{\sl Phys. Rev. } {\bf D 34}, 1967 (1986).
\bibitem{Maalampi} J. Maalampi et al.,
{\sl Nucl. Phys. } {\bf B 207}, 233 (1982).
\bibitem{Szul} N.B. Shul'gina,
{\sl Phys. Rev. Lett.} {\bf 73}, 2658 (1994).
\bibitem{Sever} N. Severijns et al.,
{\sl Nucl. Phys.} {\bf A 629}, 423c (1998).
\bibitem{Data} Review of Particle Physics, C. Caso et al.,
{\sl Eur. Phys. J.} {\bf C 3}, 1 (1998).
\bibitem{Zralek} M. Zra\l{}ek et al.,
{\sl Phys. Rev.} {\bf D  59}, 013010 (1999).
\bibitem{Olness} F. I. Olness et al.,
{\sl Phys. Rev.} {\bf D  30}, 1034 (1984).
\bibitem{Lang} P. Langacker et al.,
{\sl Phys. Rev.} {\bf D  40}, 1569 (1989).
\bibitem{Fetscher} W. Fetscher, at al.,
{\sl Phys.  Lett.} {\bf B 173}, 102 (1986).
\bibitem{Lee} T. D. Lee, C. N. Yang,
{\sl Phys. Rev.} {\bf 104}, 254 (1956).
\bibitem{Sromicki} J. Sromicki,
{\sl Institut f\"ur Teilchenphysik, ETH Z\"urich}, (1994).
\bibitem{Armbr} B. Armbruster  et al.,
{\sl Phys. Rev. Lett.}  {\bf  81}, 520 (1998).
\bibitem{MAbe} M. Abe, et al.,
{\sl Phys. Rev. Lett.} {\bf 83}, 4253 (1999).
\bibitem{Morit} M. Morita,
{\sl Beta decay and muon capture}, W. A. Benjamin, INC., Reading,
Massachusetts 1973, p. 330.
\bibitem{John} C.H. Johnson  et al.,
{\sl Phys. Rev.}  {\bf  132}, 1149 (1963).
\bibitem{Louis} E. Church  et al.,
FERMILAB-P-0898, (1997).
\bibitem{Boone} E898 Collaboration, R. Stefanski,
FERMILAB-CONF-98-141-E, (1998), Talk given at International Workshop on
JHF Science (JHF 98), Tsukuba, Japan, 4-7 Mar 1998.
\bibitem{Miniboone} A. O. Bazarko,
Talk presented at American Physical Society (APS) Meeting of the Division of
Particles and Fields (DPF 99), Los Angeles,  CA, 5-9 Jan 1999, hep-ex/9906003.
\end{thebibliography}
\end{document}